\documentclass[aps,pra,showpacs,superscriptaddress]{revtex4}
\usepackage{amsfonts}
\usepackage{amsmath}
\usepackage{amssymb}
\usepackage{graphicx}

\begin{document}

\title{Topological defects in rotating spin-orbit-coupled dipolar spin-1
Bose-Einstein condensates}
\author{Ning Su}
\affiliation{Key Laboratory for Microstructural Material Physics of Hebei Province,
School of Science, Yanshan University, Qinhuangdao 066004, China}
\author{Qingbo Wang}
\affiliation{Key Laboratory for Microstructural Material Physics of Hebei Province,
School of Science, Yanshan University, Qinhuangdao 066004, China}
\affiliation{School of Physics and Technology, Tangshan Normal University, Tangshan
063000, China}
\author{Jinguo Hu}
\affiliation{Key Laboratory for Microstructural Material Physics of Hebei Province,
School of Science, Yanshan University, Qinhuangdao 066004, China}
\author{Xianghua Su}
\affiliation{Key Laboratory for Microstructural Material Physics of Hebei Province,
School of Science, Yanshan University, Qinhuangdao 066004, China}
\author{Linghua Wen}
\email{linghuawen@ysu.edu.cn}
\affiliation{Key Laboratory for Microstructural Material Physics of Hebei Province,
School of Science, Yanshan University, Qinhuangdao 066004, China}
\date{\today }

\begin{abstract}
We consider the topological defects and spin structures of spin-1
Bose-Einstein condensates (BECs) with spin-orbit coupling (SOC) and
dipole-dipole interaction (DDI) in a rotating harmonic plus quartic trap.
The combined effects of SOC, DDI and rotation on the ground-state phases of
the system are analyzed. Our results show that for fixed rotation frequency
structural phase transitions can be achieved by adjusting the magnitudes of
the SOC and DDI. A ground-state phase diagram is given as a function of the
SOC and DDI strengths. It is shown that the system exhibits rich quantum
phases including vortex string phase with isolated density peaks (DPs),
triangular (square) vortex lattice phase with DPs, checkerboard phase, and
stripe phase with hidden vortices and antivortices. For given SOC and DDI
strengths, the system can display pentagonal vortex lattice with DPs, vortex
necklace with DPs, and exotic topological structure composed of multi-layer
visible vortex necklaces, a hidden giant vortex and hidden vortex necklaces,
depending on the rotation frequency. In addition, the system sustains
fascinating novel spin textures and skyrmion excitations, such as an
antiskyrmion pair, antiskyrmion-half-antiskyrmion (antiskyrmion-antimeron)
cluster, skyrmion-antiskyrmion lattice, skyrmion-antiskyrmion cluster,
skyrmion-antiskyrmion-meron-antimeron lattice, double-layer
half-antiskyrmion necklaces, and composite giant-antiskyrmion-antimeron
necklaces.
\end{abstract}

\maketitle

\section{Introduction}

Spin--orbit coupling (SOC) of a quantum particle has a significant influence
on many physical phenomena, such as quantum spin Hall effect, topological
insulators, and topological superconductors \cite{Zutic,Qi}. Recently, the
experimental realization of artificial SOC in both Bose-Einstein condensates
(BECs) and quantum degenerate Fermi gases provides us an ideal platform to
study exotic quantum phenomena and novel states of matter \cite%
{LinYJ,WangP,WuZ,HuangL,LiJR}. So far, a lot of experimental and theoretical
studies of spin-orbit-coupled Bose gases have been focused on the quantum
phases of pseudo-spin-1/2 BECs with SOC \cite%
{Zhai,Sinha,HuH,ZhangY,Kawakami,Qu,XuY,Stringari,Kartashov,LiX,FanZ}.
Particularly, the interplay between spin-orbit coupling, rotation and
interatomic interactions can lead to various topological excitations of the
BECs \cite{XuX,Radic,ZhouX,Aftalion,WangH,ZhangXF,LiuCF1,YangH}. Most
recently, spin-orbit-coupled spin-1 BECs of $^{87}$Rb atoms have been
realized experimentally \cite{Campbell,LuoX}, which opens another window for
the exploration of peculiar physical properties of spin-1 BECs generally
unaccessible in pseudo-spin-1/2 BECs and electronic materials due to the
competition among the spin-exchange interaction, SOC and the other
parameters \cite{WangC,Ruokokoski,LiuCF2,Kato,ChenL,Adhikari}.

On the other hand, quantum gases with magnetic dipole-dipole interaction
(DDI), especially the BECs with DDI, have also drawn much attention both
experimentally \cite{Griesmaier,Ray,Chomaz,Tanzi} and theoretically \cite%
{Lahaye,Kawaguchi,Goral,YiS1,Santos,Malet,LiuB,Cha,Borgh,Bisset} in recent
years. Relevant studies on spinor BECs with DDI have shown that the
interplay between the short-range spin-exchange interaction and the
long-range anisotropic DDI can lead to rich topological defects, spin
textures and spin dynamics \cite{Lahaye,Kawaguchi}. Therefore it is of
particular interest to investigate the combined effects of SOC and DDI on
spinor BECs, and this idea has recently attracted extensive attention \cite%
{XuY,Kato,DengY,Wilson,Gopalakrishnan,HuangC}. For instance, an experimental
scheme has been proposed by Deng \textit{et al.} to generate SOC in spin-3
Cr atoms using Raman processes \cite{DengY}. A thermodynamically stable
ground state with a meron spin configuration has been predicted in a dipolar
pseudo-spin-1/2 BEC with Rashba SOC \cite{Wilson}. In addition, Kato \textit{%
et al. }studied twisted spin vortices in a spin-1 BEC with Rashba SOC and
DDI confined in a cigar-shaped trap \cite{Kato}.

To the best of our knowledge, the existing studies of the BECs with SOC and
DDI as mentioned above refer to the nonrotating case. Considering that one
of the most striking hallmarks of a superfluid is its response to rotation,
in this paper we study the combined effects of rotation, SOC and DDI on the
ground-state structure and spin texture of a dipolar spin-1 BEC with SOC
confined in a rotating harmonic plus quartic trap (anharmonic trap). Based
on such a harmonic plus quartic trap, one can investigate the ground-state
and dynamic properties of spin-1 BEC even if the rotation frequency exceeds
the trapping frequency. For fixed rotation frequency, a phase diagram is
given with respect to the SOC strength and the DDI strength. We find that
the system sustains rich ground-state structures including vortex string
phase with density peaks (DPs), triangular (square) vortex lattice phase
with DPs, checkerboard phase, stripe phase with hidden vortices and
antivortices, multi-layer visible vortex necklaces with a hidden giant
vortex plus hidden vortex necklaces. For the case of fixed SOC and DDI
strengths, with the increase of rotation frequency the system can exhibit
pentagonal vortex lattice with DPs, vortex necklaces with DPs and a central
Mermi-Ho vortex \cite{Mermin} or with a hidden giant vortex plus hidden
vortex necklaces \cite{Wen1,Wen2}. Furthermore, the system supports exotic
spin textures and skyrmion excitations including an antiskyrmion pair,
antiskyrmion-antimeron cluster, skyrmion-antiskyrmion lattice,
skyrmion-antiskyrmion cluster, skyrmion-antiskyrmion-meron-antimeron
lattice, double-layer antimeron necklaces, and composite
giant-antiskyrmion-antimeron necklaces.

The paper is organized as follows. The theoretical model for the system is
introduced in section 2. In section 3, we present and analyze the
topological structures and typical spin textures of the system. Finally we
give a brief summary in section 4.

\section{Theoretical model}

By assuming tight confinement in the $z$-direction, we consider a rotating
quasi-two-dimensional (quasi-2D) spin-1 BEC with Rashba SOC and DDI in a
harmonic plus quartic trap. In the mean-field approximation, the dynamics of
the system obeys the coupled 2D Gross-Pitaevskii (GP) equations \cite%
{Kato,Lahaye,Kawaguchi,Kawaguchi2},%
\begin{equation}
i{\hbar }\frac{{\partial {\psi _{m}}}}{{\partial t}}{=}\left( {-\frac{{{%
\hbar ^{2}}}}{{2M}}{\nabla }^{2}+V_{tr}+c_{0}n-\Omega {L_{z}}}\right) {\psi
_{m}}+c_{1}\sum\limits_{{m^{\prime }}=-1}^{1}\mathbf{F}\cdot {\mathbf{f}_{m{%
m^{\prime }}}}{\psi _{{m^{\prime }}}}+\sum\limits_{{m^{\prime }}=-1}^{1}{{{({%
V_{so}})}_{m{m^{\prime }}}}}{\psi }_{{m^{\prime }}}+c_{dd}\sum\limits_{{%
m^{\prime }}=-1}^{1}{{\mathbf{b}}}\cdot {\mathbf{f}_{m{m^{\prime }}}}{\psi _{%
{m^{\prime }}}},  \label{GPE}
\end{equation}%
where $M$ is the atomic mass, ${\psi _{m}}(m=1,0,-1)$ is the 2D macroscopic
wave function of the atoms condensed in the spin state $\left\vert
1,m\right\rangle $, and the total atomic density $n={n_{1}}+{n_{0}}+{n_{-1}}%
=\sum_{m}{{{\left\vert {{\psi _{m}}}\right\vert }^{2}}}$ satisfies $\int nd%
\mathbf{r=}N$ with $N$ being the total number of atoms\textit{.} The
harmonic plus quartic trap ${V}_{tr}$ is given by \cite{Fetter}%
\begin{equation}
V_{tr}=\frac{1}{2}M\omega _{\bot }^{2}\left( {r}_{\bot }^{2}{+\mu \frac{{r}%
_{\bot }^{4}}{{a^{2}}}}\right) =\frac{1}{2}\hbar \omega _{\bot }\left( {%
\frac{{r}_{\bot }^{2}}{{a^{2}}}+\mu \frac{{r}_{\bot }^{4}}{{a^{4}}}}\right) ,
\label{trap}
\end{equation}%
where $\omega _{\perp }$ is the radial trap frequency, $r_{\bot }=\sqrt{%
x^{2}+y^{2}}$, $a=\sqrt{\hbar /M\omega _{\perp }}$, and $\mu $ is a
dimensionless constant which denotes the anharmonicity of the trap. $\Omega $
is the rotation frequency along the $z$-direction, and $L_{z}=i\hbar \left(
y\partial _{x}-x\partial _{y}\right) $ is the $z$ component of the
angular-momentum operator. The Rashba SOC is expressed by ${V_{{so}}}%
=k\left( {{{f}_{x}p}{_{x}}+{f_{y}p}{_{y}}}\right) $, where $k$ represents
the isotropic SOC strength \cite{XuZF,LiJ}, and ${p_{\alpha }}=-i\hbar
\partial _{\alpha }(\alpha =x,y)$ is the momentum operator. ${c_{0}}=4\pi {%
\hbar ^{2}}(2{a_{2}}+{a_{0}})\gamma /3M$ and ${c_{1}}=4\pi {\hbar ^{2}}({%
a_{2}}-{a_{0}})\gamma /3M$ give the strengths of density-density and
spin-exchange interactions, respectively. Here $\gamma =\sqrt{{M{\omega _{z}/%
}2\pi \hbar }}$ with ${{\omega _{z}}}$ being the harmonic trap frequency in
the $z$-direction, and ${a_{s}}(s=0,2)$ is the $s$-wave scattering length
for the scattering channel with total spin $s$. $\boldsymbol{F}=\left(
F_{x},F_{y},F_{z}\right) $ is the spin vector defined by ${F_{\mathcal{\nu }}%
}(\mathbf{r})\equiv \sum_{mm^{\prime }}{\psi _{m}^{\ast }}{\left( {{f_{%
\mathcal{\nu }}}}\right) _{m{m^{\prime }}}}{\psi _{{m^{\prime }}}}$ $(%
\mathcal{\nu }=x,y,z)$, and $\boldsymbol{f}$ $=(f_{x},f_{y},f_{z})$ is the
vector of $3\times 3$ spin-1 matrices. Concretely, the components of the
spin vector $\boldsymbol{F}$ are expressed as
\begin{eqnarray}
{F_{x}} &=&\frac{1}{\sqrt{2}}\left[ {\psi _{1}^{\ast }{\psi _{0}}+\psi
_{0}^{\ast }\left( {{\psi _{1}}+{\psi _{-1}}}\right) +\psi _{-1}^{\ast }{%
\psi _{0}}}\right] ,  \label{Fx} \\
{F_{y}} &=&\frac{i}{\sqrt{2}}\left[ {-\psi _{1}^{\ast }{\psi _{0}}+\psi
_{0}^{\ast }\left( {{\psi _{1}}-{\psi _{-1}}}\right) +\psi _{-1}^{\ast }{%
\psi _{0}}}\right] ,  \label{Fy} \\
{F_{z}} &=&{\left\vert {{\psi _{1}}}\right\vert ^{2}}-{\left\vert {{\psi
_{-1}}}\right\vert ^{2}.}  \label{Fz}
\end{eqnarray}

The last term on the right-hand side of equation (\ref{GPE}) corresponds to
the magnetic DDI, where ${c_{dd}}={\mu _{0}}g_{e}^{2}\mu _{B}^{2}\gamma
/4\pi $ denotes the DDI coefficient, ${\mu _{0}}$ is the vacuum magnetic
permeability, $\mu _{B}$ is the Bohr magneton, and $g_{e}$ is the Land\'{e}
g-factor. Here ${{\mathbf{b}}}$ is the effective 2D dipole field defined by $%
b_{\nu }\equiv \int d\mathbf{r}_{\bot }^{\prime }\sum_{\nu \nu ^{\prime
}}Q_{\nu \nu {^{\prime }}}^{2D}{F_{\nu {^{\prime }}}}$ $(\mathcal{\nu },%
\mathcal{\nu }^{\prime }=x,y,z)$ in which $Q_{\nu \nu {^{\prime }}}^{2D}$ is
the 2D dipole kernel related to the 3D one. The 3D dipole kernel is given by
\begin{equation}
{Q_{\nu \nu {^{\prime }}}^{3D}}\left( {\mathbf{r}-{\mathbf{r}^{\prime }}}%
\right) =\frac{{{\delta _{\nu {\nu }^{\prime }}}-3{{\hat{r}}_{\nu }}{{\hat{r}%
}}}_{\nu ^{\prime }}}{{{r^{3}}}},  \label{3DDipoleKernel}
\end{equation}%
where $\mathbf{r=(}x,y,z\mathbf{)}$\textbf{, }$r=|\mathbf{r}|$, $\hat{r}=%
\mathbf{r}/r$, and $\left\vert {\mathbf{r}-{\mathbf{r}^{\prime }}}%
\right\vert $ is the relative position of two dipoles. As the motion of the
BEC in the $z$-direction is frozen in the ground state of harmonic
oscillator, we assume a Gaussian profile along the $z$-direction, i.e., $%
\varphi _{m}(z)=(1/\pi ^{1/4}a_{z}^{1/2})\exp (-z^{2}/2a_{z}^{2})$ with ${{%
a_{z}=}}\sqrt{\hbar /M\omega _{z}}$, then we can reduce ${Q_{\nu \nu {%
^{\prime }}}^{3D}}$ to $Q_{\nu \nu {^{\prime }}}^{2D}$ in two dimensions.
The detailed derivation is referred to the relevant literatures \cite%
{Lahaye,Kawaguchi,Kawaguchi2}. For the sake of numerical calculation, we use
the Fourier transform and convolution theorem to express the 2D dipole
kernel and the 2D effective dipole field in momentum space, and we obtain%
\begin{equation}
\tilde{Q}^{2D}\left( k_{x},k_{y}\right) =-\frac{{4\pi }}{3}\left(
\begin{array}{ccc}
1 & 0 & 0 \\
0 & 1 & 0 \\
0 & 0 & -2%
\end{array}%
\right) +4\pi \frac{{G\left( {\frac{{{k_{\bot }}{a_{z}}}}{\sqrt{2}}}\right) }%
}{{k_{\bot }^{2}}}\left(
\begin{array}{ccc}
{k_{x}^{2}} & {{k_{x}}{k_{y}}} & 0 \\
{{k_{x}}{k_{y}}} & {k_{y}^{2}} & 0 \\
0 & 0 & {-k_{\bot }^{2}}%
\end{array}%
\right) ,  \label{Fourier2DDipoleKernel}
\end{equation}%
\begin{equation}
b_{v}={\mathcal{F}^{-1}}\left\{ {\sum\limits_{{\nu }^{\prime }}{\tilde{Q}%
^{2D}}\left( k_{x},k_{y}\right) \mathcal{F}\left[ {{F_{{\nu }^{\prime }}}}%
\right] }\right\} ,  \label{Fourier2DDipoleField}
\end{equation}%
where ${k_{\bot }}=\sqrt{k_{x}^{2}+k_{y}^{2}}$, $G(q)\equiv
2qe^{q^{2}}\int_{q}^{\infty }e^{-t^{2}}dt$, and ${\mathcal{F}}$ represents
the Fourier transform operator \cite{Kawaguchi,Kawaguchi2}.

In our calculation, it is convenient to introduce the following notations $%
\widetilde{r}_{\bot }=r_{\bot }/a$, $\widetilde{t}=\omega _{\perp }t$, $%
\widetilde{V}_{tr}=V_{tr}/\hbar \omega _{\perp }=(\widetilde{r}^{2}+\mu
\widetilde{r}^{4})/2$,$\ \widetilde{\Omega }=$ $\Omega /\omega _{\perp }$, $%
\widetilde{L}_{z}=L_{z}/\hbar $, $\widetilde{\psi }_{m}=\psi _{m}a/\sqrt{N}($
$m=1,0,-1)$, $\widetilde{c}_{0}=c_{0}N/\hbar \omega _{\perp }a^{2}$, $%
\widetilde{c}_{1}=c_{1}N/\hbar \omega _{\perp }a^{2}$, $\widetilde{k}%
=k/\omega _{\perp }a$, and $\widetilde{c}_{dd}=c_{dd}N/\hbar \omega _{\perp
}a^{2}$. Then we obtain the dimensionless coupled GP equations%
\begin{equation}
i\frac{{\partial {\psi _{m}}}}{{\partial t}}{=}\left( {-\frac{{1}}{{2}}{%
\nabla }^{2}+V_{tr}+c_{0}n-\Omega {L_{z}}}\right) {\psi _{m}}%
+c_{1}\sum\limits_{{m^{\prime }}=-1}^{1}\mathbf{F}\cdot {\mathbf{f}_{m{%
m^{\prime }}}}{\psi _{{m^{\prime }}}}+\sum\limits_{{m^{\prime }}=-1}^{1}{{{({%
V_{so}})}_{m{m^{\prime }}}}}{\psi }_{{m^{\prime }}}+c_{dd}\sum\limits_{{%
m^{\prime }}=-1}^{1}{{\mathbf{b}}}\cdot {\mathbf{f}_{m{m^{\prime }}}}{\psi _{%
{m^{\prime }}}.}  \label{DimensionlessGPE}
\end{equation}%
For simplicity the tilde is omitted throughout this paper.

For spin-1 BEC, the spin texture is defined by \cite{Mizushima,Kasamatsu}%
\begin{eqnarray}
{S_{x}} &=&\frac{1}{\sqrt{2}}\frac{{\psi _{0}^{\ast }{\psi _{1}}+(\psi
_{1}^{\ast }+\psi _{-1}^{\ast }){\psi _{0}}+\psi _{0}^{\ast }{\psi _{-1}}}}{{%
{{\left\vert {{\psi _{1}}}\right\vert }^{2}}+{{\left\vert {{\psi _{0}}}%
\right\vert }^{2}}+{{\left\vert {{\psi _{-1}}}\right\vert }^{2}}}},
\label{SpinDensityX} \\
{S_{y}} &=&\frac{i}{\sqrt{2}}\frac{{\psi _{0}^{\ast }{\psi _{1}}+(\psi
_{-1}^{\ast }-\psi _{1}^{\ast }){\psi _{0}}-\psi _{0}^{\ast }{\psi _{-1}}}}{{%
{{\left\vert {{\psi _{1}}}\right\vert }^{2}}+{{\left\vert {{\psi _{0}}}%
\right\vert }^{2}}+{{\left\vert {{\psi _{-1}}}\right\vert }^{2}}}},
\label{SpinDensityY} \\
{S_{z}} &=&\frac{{\psi _{1}^{\ast }{\psi _{1}}-\psi _{-1}^{\ast }{\psi _{-1}}%
}}{{{{\left\vert {{\psi _{1}}}\right\vert }^{2}}+{{\left\vert {{\psi _{0}}}%
\right\vert }^{2}}+{{\left\vert {{\psi _{-1}}}\right\vert }^{2}}}}.
\label{SpinDensityZ}
\end{eqnarray}%
The spatial distribution of the topological structure of the system can be
well described by the topological charge density%
\begin{equation}
q(x,y)=\frac{1}{{4\pi }}\mathbf{s}\cdot \left( {\frac{{\partial \mathbf{s}}}{%
{\partial x}}\times \frac{{\partial \mathbf{s}}}{{\partial y}}}\right) ,
\label{topological charge density}
\end{equation}%
with $\mathbf{s=S/}\left\vert \mathbf{S}\right\vert $, and the topological
charge is defined as%
\begin{equation}
Q=\int {q(x,y)dxdy.}  \label{topological charge Q}
\end{equation}%
The topological charge $\left\vert Q\right\vert $ is unchanged, no matter
how one exchange the components of the spin density vector ${S_{x}}$, ${S_{y}%
}$ and ${S_{z}}$,

\section{Results and discussion}

Due to the presence of SOC, rotation, nonlocal DDI and nonlinear contact
interaction, there is no analytical solution for the coupled nonlinear GP
equations (\ref{GPE}) and (\ref{DimensionlessGPE}). Next, we numerically
solve the 2D GP equations (\ref{DimensionlessGPE}) to obtain the ground
state by minimizing the total energy of the system using the widely adopted
imaginary-time propagation method \cite{ZhangY,Wen1,Wen2}. A remarkable
feature of this system is that there are a large number of free parameters,
including the rotation frequency, the strength and sign of the
density-density interaction and spin-exchange interaction, SOC and DDI, and
the anharmonicity of the trap. As a result of the competition among multiple
parameters, the system can exhibit rich ground-state structures and unique
properties. In order to highlight the effects of the SOC, DDI and rotation,
without loss of generality, we fix $\mu =0.5$, $c_{0}=1000$ and $c_{1}=50$.
For other values of $\mu $, $c_{0}$ and $c_{1}$, our simulation shows that
there exist similar phase diagrams and ground-states configurations.

\subsection{SOC and DDI effects}

In this system, rotation effect is dominant compared with SOC and DDI
effects. Therefore, in this section we choose a relatively small rotation
frequency (for instance, $\Omega =0.1$ or $\Omega =0.4$), which can not only
reflect the combined effects of SOC, DDI and rotation, but also not
highlight the rotation effect. In the case of fixed rotation frequency, we
give a ground-state phase diagram spanned by the SOC strength $k$ and the
DDI strength ${c_{dd}}$. There are five different phases marked by A$-$E,
which differs in terms of their density profiles and phase distributions. In
the following discussion, we will give a detailed description of each phase.
The density and phase distributions of the five different phases A$-$E in
figure 1 are shown in figures 2(a)--2(e), respectively. In figure 2, the
left three columns are the density profiles $\left\vert \psi _{1}\right\vert
^{2}$, $\left\vert \psi _{0}\right\vert ^{2}$ and $\left\vert \psi
_{-1}\right\vert ^{2}$ of three components $m_{F}=1$ (component 1), $m_{F}=0$
(component 2) and $m_{F}=-1$ (component 3), columns 4 to 6 denote the
corresponding phase distributions $\theta _{1}=\arg \psi _{1}$, $\theta
_{0}=\arg \psi _{0}$ and $\theta _{-1}=\arg \psi _{-1}$, respectively, and
the last column represents the total density $n$.

\begin{figure}[tbh]
\centerline{\includegraphics*[width=7cm]{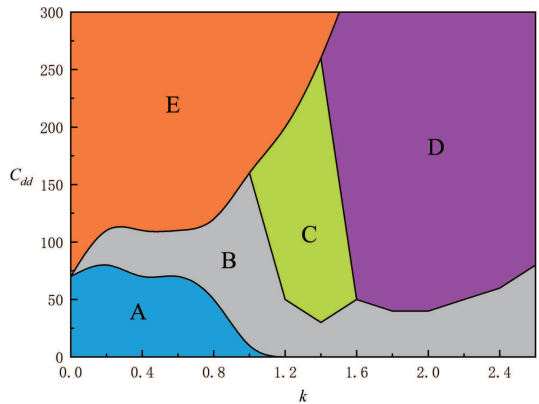}}
\caption{Ground-state phase diagram of rotating spin-orbit-coupled dipolar
spin-1 BEC with respect to $k$ and ${c_{dd}}$ for $c_{0}=1000$, $c_{1}=50$
and $\Omega =0.1$. There are five different phases marked by A$-$E.}
\label{Figure1}
\end{figure}

We start from the case where the SOC and DDI are weak, which is indicated by
the region A in figure 1. In this phase, there exist a hidden vortex \cite%
{Wen1,Wen2,Wen3}, a vortex pair or a vortex string in the three components,
where the density distribution of $m_{F}=1$ component forms two petal-like
density peaks which are very similar to two quantum droplets \cite%
{Ferrier,Kartashov2}, and the phase defects in the components are basically
arranged along the same straight line. To distinguish it from other phases,
we may call it vortex string phase with DPs. With the increase of SOC or
DDI, the A phase transform to the B phase, as shown in figure 1. The density
peaks in the three components show evident triangular distribution, with an
extra droplet, a vortex or a vortex pair being located in the central region
of the trap (see figure 2(b)). At the same time, the vortices except the
central ones in the components constitute a triangular vortex lattice. This
phase is a typical triangular vortex lattice phase with DPs. In a narrow
parameter regime of relatively large SOC strength and DDI strength, the C
phase emerges as the ground state. Typical density and phase distributions
are given in figure 2(c). The trap center is occupied by a droplet, a vortex
and a doubly quantized vortex for the three components, respectively. The
main density peaks are arranged in a square. In this context, the C phase is
a square vortex lattice phase with DPs.\ For the case of strong SOC, with
the increase of ${c_{dd}}$, the B phase transforms to the D phase
(checkerboard phase), where the component densities evolve into checkerboard
patterns, the vortices in $m_{F}=1$ and $m_{F}=-1$ components disappear, and
an alternative vortex-antivortex pair lattice is generated in $m_{F}=0$
component (see figure 2(d)). The phenomenon of component separation is
significantly enhanced in\ the checkerboard phase.

\begin{figure}[tbph]
\centerline{\includegraphics*[width=10cm]{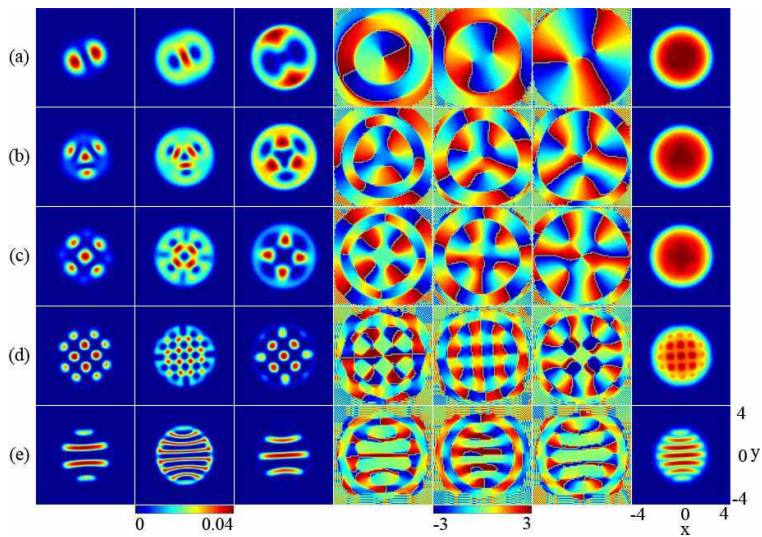}}
\caption{Typical ground states of spin-1 BECs with SOC and DDI in a rotating
harmonic plus quartic trap, where $\Omega =0.1$. Rows (a)$-$(e) correspond
to the A$-$E phases in figure 1, respectively. (a) $k=0.8$, ${c_{dd}=40}$,
(b) $k=1.4$, ${c_{dd}=20}$, (c) $k=1.4$, ${c_{dd}=50}$, (d) $k=2$, ${%
c_{dd}=300}$, and (e) $k=0.8$, ${c_{dd}=300}$. The columns from left to
right denote ${{{\left\vert {{\protect\psi _{1}}}\right\vert }^{2}}}$, ${{{%
\left\vert {\protect\psi }_{0}\right\vert }^{2}}}$, ${{{\left\vert {{\protect%
\psi _{-1}}}\right\vert }^{2}}}$, arg${{{{{\protect\psi _{1}}}}}}$, arg${{{{{%
\protect\psi _{0}}}}}}$, arg${{{{{\protect\psi _{-1}}}}}}${,} and the total
density $n=\sum_{m}{{{\left\vert {{\protect\psi _{m}}}\right\vert }^{2}}}%
(m=1,0,-1)$, respectively. The unit length is $a$.}
\label{Figure2}
\end{figure}

In the limit of strong DDI but weak SOC, the system sustains E phase.
Typical density and phase distributions of such a phase are shown in figure
2(e), where stripe density patterns are created in individual components,
and there are only few hidden vortices plus hidden antivortices, and some
ghost vortices \cite{Wen1,Tsubota} as well as ghost antivortices distributed
in outskirts of the atom cloud. Thus the E phase is a stripe phase. In
addition, for the A$-$C phases, the total densities show typical
Thomas-Fermi (TF) distribution. By contrast, for the D phase and the E
phase, the total densities exhibit checkerboard pattern and stripe pattern,
respectively. Note that here the stripe phase (E phase) is remarkably
different from the irregular stripe phase reported in a spin-orbit-coupled
dipolar spin-1 BEC without spin-exchange interaction in a cigar-shaped 3D
harmonic trap \cite{Kato}. The latter case exists for strong SOC but with
weak DDI, where each component density shows no symmetry with respect to the
two principal axes of the atom cloud, the total density exhibits TF
distribution, and the spin texture has a spin stripe structure (see figure
2(a) in Ref. \cite{Kato}). However, in the present system, the stripe state
exists for strong DDI but with weak or relatively weak SOC, where both the
component densities and the total density exhibit good symmetry concerning
the principal axes of the atom cloud, and the spin texture is an exotic
skyrmion-antiskyrmion-half-skyrmion-halt-antiskyrmion lattice (see figure
6(d)). This point is due to the complex competition between SOC, DDI,
rotation, spin-exchange interaction and density-density interaction.

To analyze and compare the effects of SOC and DDI on the ground-state of the
system, in figure 3 we show the density and phase distributions of spin-1
BECs with various SOC strengths or DDI strengths in a rotating anharmonic
trap. For weak SOC ($k=0.8$), the system shows a triangular vortex lattice
structure with isolated DPs (see figure 3(a)), where for component 1 there
is a vortex at the trap center surrounded by several ghost vortices, a
visible triangular vortex lattice and a central antivortex are generated in
component 2 with several ghost vortices and antivortices being distributed
in the outskirts of the cloud, and a hidden triangular vortex lattice is
formed in component 3. When $k$ increases to $1.2$ (figure 3(b)), there are
four obvious DPs in component 1 with a doubly quantized vortex at the trap
center, and a square vortex lattice is formed in component 3. At the same
time, a square vortex lattice and a central antivortex constitute a visible
criss-cross vortex string lattice in component 2. Thus we may call it square
vortex lattice with DPs. With the further increase of $k$, e.g., $k=2.4$,
the ground state of the system becomes a checkerboard phase (figure 3(c)),
which is similar to that in Fig. 2(e). From figures 3(a)-3(c), it is shown
that the increase of SOC strength can lead to a structural phase transition
from a triangular vortex lattice with DPs to a square vortex lattice with
DPs and then to a checkerboard phase. Physically, this phenomenon is induced
by the spin-orbit interaction between the atomic spin and the center-of-mass
motion of the BEC. Therefore varying the SOC strength will result in the
change of the atomic spin structure and the spin texture of the system. When
the SOC strength increases, the DP number and the vortex (or antivortex)
number in each component evidently increases because the stronger SOC means
there is a larger orbital angular momentum and more energy input into the
system \cite{Zhai,XuX}. On the other hand, the interplay among SOC, DDI,
rotation and contact interaction will change the symmetry of vortex
distribution in the system. Consequently, for strong SOC, the checkerboard
phase rather than a triangular vortex lattice phase, dominates the
topological structure of the system.

\begin{figure}[tbh]
\centerline{\includegraphics*[width=10cm]{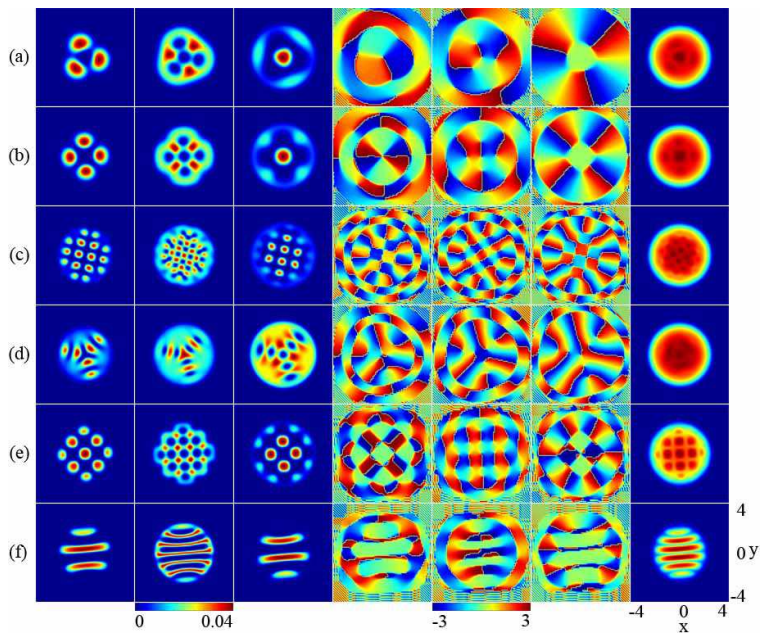}}
\caption{Ground-state density distributions and phase distributions of
spin-1 BECs with SOC and DDI in a rotating harmonic plus quartic trap. The
top three rows: ${c_{dd}=100}$, $\Omega =0.1$, (a) $k=0.8$, (b) $k=1.2$, and
(c) $k=2.4$. The bottom three rows: $k{=1.8}$, $\Omega =0.4$, (d) ${c_{dd}}%
=1 $, (e) ${c_{dd}}=300$, and (f) ${c_{dd}}=800$. The other parameters are
the same with those in figures 1 and 2. The columns from left to right
denote ${{{\left\vert {{\protect\psi _{1}}}\right\vert }^{2}}}$, ${{{%
\left\vert {\protect\psi }_{0}\right\vert }^{2}}}$, ${{{\left\vert {{\protect%
\psi _{-1}}}\right\vert }^{2}}}$, arg${{{{{\protect\psi _{1}}}}}}$, arg${{{{{%
\protect\psi _{0}}}}}}$, arg${{{{{\protect\psi _{-1}}}}}}${,} and the total
density $n$, respectively. The unit length is $a$.}
\label{Figure3}
\end{figure}

Figures 3(d)-3(f) illustrate the effect of DDI on the ground-state structure
of the system, where $k=1.8$ and $\Omega =0.4$. For weak DDI, e.g., ${c_{dd}}%
=1$, the ground state displays obvious vortex chain structure in which three
vortex chains in each component form an angle of $2\pi /3$ degrees with each
other as shown in figure 3(d). When the DDI strength increases to ${c_{dd}}%
=300$, the checkerboard phase becomes the ground-state phase of the system
(see figure 3(e)). For the larger DDI strength ${c_{dd}}=800$, the
checkerboard phase transforms into a tripe phase as displayed in figure
3(f). Different from figures 2(d)-2(e) and 3(a)-3(c), the structural phase
transition from figure 3(d) to figure 3(f) and then to figure 3(e) is mainly
dominated by DDI.

\begin{figure}[tbh]
\centerline{\includegraphics*[width=7.5cm]{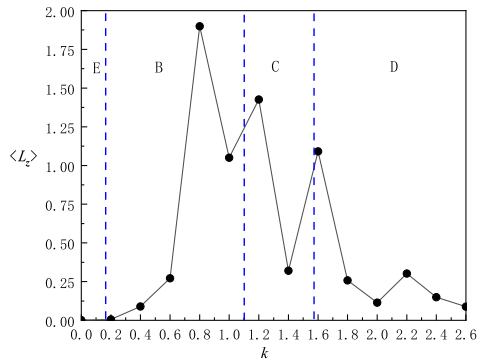}}
\caption{Orbital angular momentum $\left\langle L_{z}\right\rangle $ as a
function of $k$ for ${c_{dd}=100}$ and $\Omega =0.1$. The vertical dotted
lines separate the phases and the solid curve demonstrates the trend.}
\label{Figure4}
\end{figure}

The $z$ component of the orbital angular momentum, $\left\langle
L_{z}\right\rangle =\int \mathbf{\psi }^{\dagger }\left( \mathbf{r}\right)
\left( xp_{y}-yp_{x}\right) \mathbf{\psi }\left( \mathbf{r}\right) d\mathbf{r%
}$ with $\mathbf{\psi }\left( \mathbf{r}\right) =\left( {{\psi _{1,}\psi
_{0},\psi _{-1}}}\right) ^{\text{T}}$, is important in considering this
system. Figure 4 displays $\left\langle L_{z}\right\rangle $ as a function
of $k$ for ${c_{dd}=100}$ and $\Omega =0.1$, where B$-$E correspond to the B$%
-$E phases in figure 1, respectively. The E phase (i.e., the stripe phase)
scarcely has angular momentum because there are only few hidden vortices and
antivortices, and ghost vortices plus antivortices. Especially, the ghost
vortices (or antivortices) neither carry angular momentum nor energy. The
first rapid increase in $\left\langle L_{z}\right\rangle $ occurs in the B
phase ($0.18\lesssim k\lesssim 0.8$) due to there being typical triangular
vortex lattice, $\left\langle L_{z}\right\rangle $ reaches the maximum at $%
k\simeq 0.8$ and then decreases significantly at $k\simeq 1.0$. With an
increase of SOC strength, the ground state transforms to the C phase and the
angular momentum increases suddenly. The change trend of $\left\langle
L_{z}\right\rangle $ in the C phase is similar to that in the B phase.
Although $\left\langle L_{z}\right\rangle $ fluctuates in the D phase, which
is similar to the B and C phases, on average, the angular momentum is
obviously decreasing. The main reason is that for the checkerboard phase
there is a vortex-antivortex pair lattice in the $m_{F}=0$ component and
there are no vortices in the $m_{F}=1$ and $m_{F}=-1$ components. From
figure 4, the curve of $\left\langle L_{z}\right\rangle $ and $k$ shows no
linear correlation due to the complicated competition among the anisotropic
DDI, SOC, rotation, and nonlinear contact interaction.

\subsection{Rotation effect}

Next, we consider the influence of rotation on the ground state of the
system with given DDI strength and SOC strength. In figure 5, we show the
density distributions, the phase distributions and the total densities for
the ground states of spin-1 BECs with DDI and SOC in a rotating anharmonic
trap. Here $k=1.4$, ${c_{dd}=50}$, and rotation frequencies in rows (a)$-$%
(d) are $\Omega =0.4$, $1.2$, $1.4$, and $3.0$, respectively. For relatively
small rotation frequency $\Omega =0.4$, the density and phase profiles of
three components exhibit obvious differences (see figure 5(a)). The $m_{F}=1$
component forms a petal-like structure with a central density droplet, and
there is a hidden pentagonal vortex lattice. By comparison, a visible
pentagonal vortex lattice exists in $m_{F}=0$ and $m_{F}=-1$ components, and
there is an additional central density hole for $m_{F}=0$ component and $%
m_{F}=-1$ component, respectively. The two central density holes in $m_{F}=0$
and $m_{F}=-1$ components correspond to a singly quantized vortex encircled
by pentagonal density peaks and a doubly quantized vortex surrounded by
pentagram density peaks, respectively.

\begin{figure}[tbp]
\centerline{\includegraphics*[width=11cm]{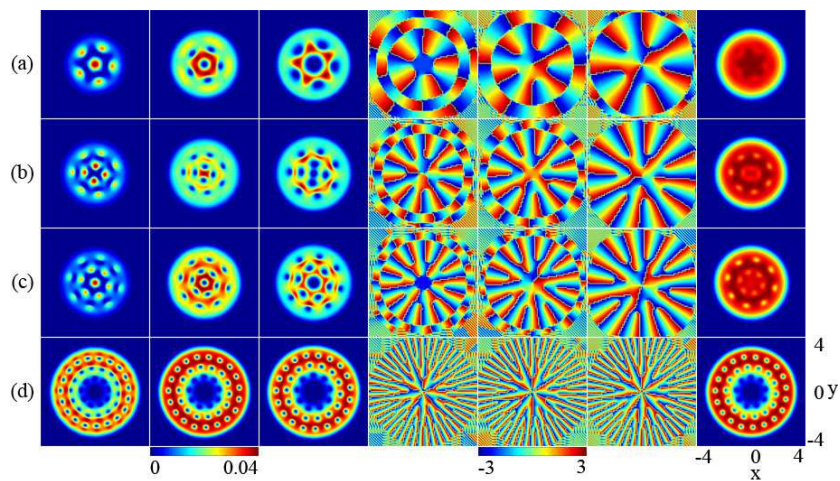}}
\caption{Ground states of spin-1 BECs with SOC and DDI in a rotating
harmonic plus quartic trap, where $c_{0}=1000$, $c_{1}=50$, $k=1.4$, and ${%
c_{dd}=50}$. (a) $\Omega =0.4$, (b) $\Omega =1.2$, (c) $\Omega =1.4$, and
(d) $\Omega =3$. The columns from left to right denote ${{{\left\vert {{%
\protect\psi _{1}}}\right\vert }^{2}}}$, ${{{\left\vert {\protect\psi }%
_{0}\right\vert }^{2}}}$, ${{{\left\vert {{\protect\psi _{-1}}}\right\vert }%
^{2}}}$, arg${{{{{\protect\psi _{1}}}}}}$, arg${{{{{\protect\psi _{0}}}}}}$,
arg${{{{{\protect\psi _{-1}}}}}}$ and the total density $n$, respectively.
The unit length is $a$.}
\label{Figure5}
\end{figure}

When the rotation frequency increases to $\Omega =1.2$, more density petals
are generated along the azimuthal direction in the $m_{F}=1$ component, and
the non-central vortices in each component are distributed along a circle,
as shown in figure 5(b). At the same time, the central regions of individual
components are occupied by a triangular vortex lattice, a density peak, and
criss-cross vortex trains, respectively. Our numerical results show that
with the increase of rotation frequency not only the number of vortices
along a single circle but also the number of circles increases, as displayed
in figures 5(b) and 5(c). In particular, for large rotation frequency, e.g.,
$\Omega =3$, the system exhibits an interesting and unique ground-state
structure as shown in figure 5(d). There is an almost full overlap of the
density profiles and the phase profiles of the three components. Here the
visible vortices form multi-layer ringlike structures, i.e., multi-layer
visible vortex necklaces. Our numerical simulation shows that the region of
the large density hole in each component is occupied by several hidden
vortex necklaces and a central hidden giant vortex (see figure 5(d)), which
is quite different from the common prediction results in rotating
spin-orbit-coupled BECs in a harmonic trap \cite%
{XuX,Radic,ZhouX,Aftalion,LiuCF1}. For the latter case, the large density
hole in the density profile at large rotation frequency characterizes a pure
giant vortex. In addition, the phase mixing of the three components can be
understood. Physically, the rapid rotation or the strong SOC will lead to
large kinetic energy. Here, the kinetic energy acts against the interspecies
contact interaction and DDI. The latter is responsible to component
separation while the former tends to expand the BECs and therefore sustains
component mixing. At the same time, the anharmonic external potential tends
to trap the BECs more tightly and hence also favors component mixing.
Consequently, component demixing can be suppressed by the kinetic energy and
external trapping potential in some conditions as shown in figure 5(d).
Essentially, the topological states of the system in figures 5(a)-5(c) are
coreless vortex states because there are no singular points in the total
density of the system while that in figure 5(d) corresponds to a nucleated
vortex state as there are obvious singular points in the total density,
which indicates the increase in rotation frequency can lead to a series of
structural phase transitions. Furthermore, we can conclude that for the case
of large $\Omega $ the rotation effect plays a key role in determining the
ground-state phase of the system compared with the DDI effect and the SOC
effect.

Based on the above discussion, as three new degrees of freedom, SOC, DDI and
rotation frequency can be used to achieve the desired ground-state phases
and to control the phase transitions between various ground states of spin-1
BECs in a harmonic plus quartic trap.

\subsection{Topological charge densities and spin textures}

To further elucidate the ground-state properties, we now analyze the
topological charge densities and spin texture of the system. In figure 6, we
show the topological charge densities (column 1) and the spin textures
(column 2) of the system in the left two columns, and the local
amplifications of the spin textures are displayed in the right two columns.
The ground states for figures 6(a)-6(d) are given in figures 2(a), 2(b),
2(d), and 2(e), respectively. Shown in figure 7 are the topological charge
densities, the spin textures, and the local enlargements of the spin
textures for the parameters in figures 3(a), 3(c), 5(b), and 5(c). For the
sake of discussion, we use red square (or rectangle), red circle, blue
triangle, blue hexagon, and blue diamond to denote a skyrmion \cite{Skyrme},
a half-skyrmion (meron) \cite{Mermin}, an antiskyrmion, a half-antiskyrmion
(antimeron), and a giant antiskyrmion, respectively.

\begin{figure}[tbph]
\centerline{\includegraphics*[width=12cm]{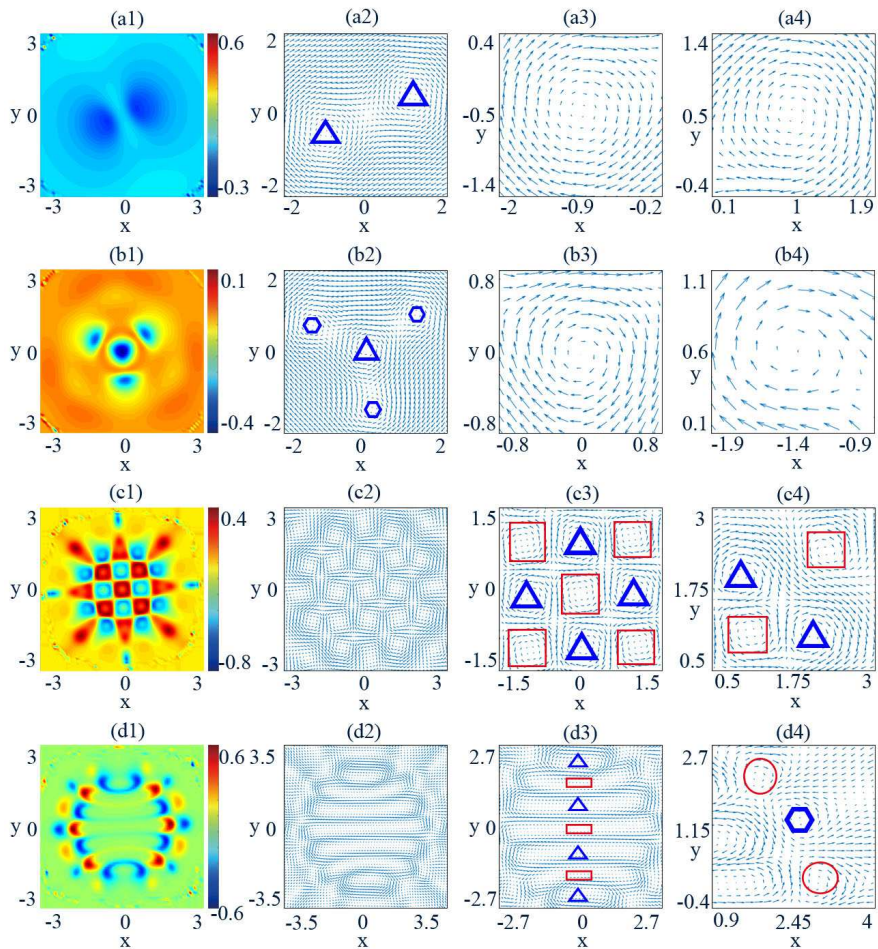}}
\caption{Topological charge densities and spin textures of rotating spin-1
BECs with DDI and SOC in a harmonic plus quartic trap, where the
corresponding ground states for (a)-(d) are given in figures 2(a), 2(b),
2(d), and 2(e), respectively. The first column (from left to right)
represents topological charge density, the second column is spin texture,
and the right two columns denote the local amplifications of the spin
texture. The red square (or rectangle), red circle, blue triangle, and blue
hexagon denote a skyrmion, a half-skyrmion, an antiskyrmion, and a
half-antiskyrmion, respectively. The unit length is $a$.}
\end{figure}

Our computation results show that the local topological charges in figures
6(a3) and 6(a4) approach $Q=-1$, which indicates that the local topological
defects in figures 6(a3) and 6(a4) are circular antiskyrmions. Thus the spin
texture in figure 6(a2) forms a fascinating antiskyrmion pair. The spin
defect in figure 6(b2) is an antiskyrmion-half-antiskyrmion
(antiskyrmion-antimeron) cluster \cite{LiuCF1,Mermin,Kasamatsu,Skyrme},
where the two local enlargements in figures 6(b3) and 6(b4) correspond to an
antiskyrmion and a half-antiskyrmion (antimeron), respectively. In
particular, the topological charge density in figure 6(c1) shows a
checkerboard pattern, and the spin texture is displayed in figure 6(c2),
where the local amplifications of the spin texture are exhibited in figures
6(c3) and 6(c4). It is shown that the topological structure in figure 6(c2)
is an interesting interlaced skyrmion-antiskyrmion lattice consists of
skyrmions with unit topological charge $Q=1$ and antiskyrmions with
topological charge $Q=-1$. From figure 6(d3), we can see that there is a
staggered skyrmion-antiskyrmion chain along the $y$ direction. At the same
time, the local topological defects in figure 6(d4) are three interlaced
half-skyrmions (merons) and half-antiskyrmion (antimeron). Thus the
topological structure in figure 6(d2) is an exotic complex
skyrmion-antiskyrmion-meron-antimeron lattice composed of a staggered
skyrmion-antiskyrmion chain and an outer circular interlaced meron-antimeron
string.

\begin{figure}[tbph]
\centerline{\includegraphics*[width=12cm]{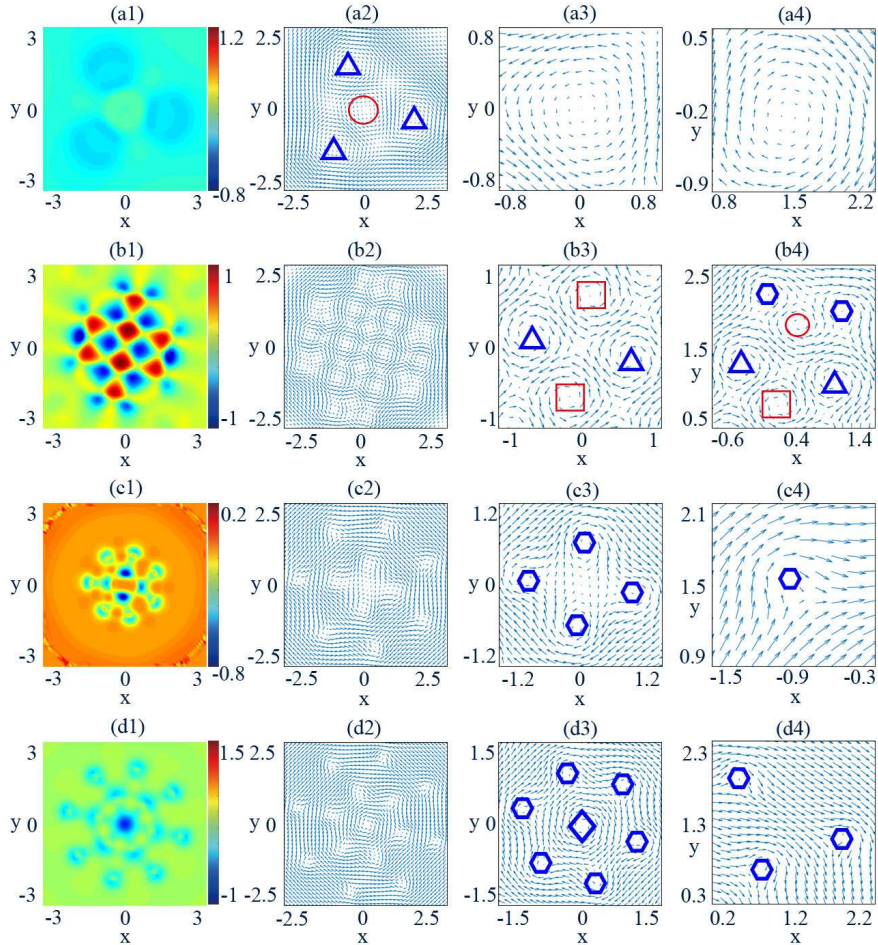}}
\caption{Topological charge densities and spin textures of rotating spin-1
BECs with DDI and SOC in a harmonic plus quartic trap, where the
corresponding ground states for (a)-(d) are given in figures 3(a), 3(c),
5(b), and 5(c), respectively. The first column represents topological charge
density, the second column is spin texture, and the right two columns denote
the local amplifications of the spin texture. The red square, red circle,
blue triangle, blue hexagon, and blue diamond represent a skyrmion, a
half-skyrmion, an antiskyrmion, a half-antiskyrmion, and a giant
antiskyrmion, respectively. The unit length is $a$. }
\end{figure}

Finally, the topological charge densities and spin textures in figure 7 are
richer and more interesting, where the ground states are displayed in
figures 3(a), 3(c), 5(b), and 5(c). Our numerical calculation shows that the
local topological charges in figures 7(a3) and 7(a4) are $Q=1$ and $Q=-1$,
respectively. Therefore the spin texture in figure 7(a2) is a special
skyrmion-antiskyrmion cluster composed of a central skyrmion and an outer
triangular antiskyrmion lattice. The topological charge density in figure
7(b1) exhibits an obvious checkerboard pattern, which is somewhat similar to
that in figure 6(c1). However, our simulation demonstrates that the
topological structure in figure 7(b2) is quite different from that in figure
6(c2). As a matter of fact, the local topological defects in figure 7(b3)
are alternating skyrmions and antiskyrmions, and the those in figure 7(b4)
include a skyrmion, two antiskyrmions, a half-skyrmion (meron) with
topological charge $Q=0.5$, and two half-antiskyrmions (antimerons) with
respective topological charge $Q=-0.5$, which indicates the spin defects in
figure 7(b2) constitute a rather complicated
skyrmion-antiskyrmion-meron-antimeron lattice. From figures 7(c2)-7(c4), one
can see that the topological structure in figure 7(c2) is peculiar
double-layer and concentric half-antiskyrmion necklaces. Particularly, the
topological charge density in figure 7(d1) has excellent discrete rotational
symmetry, which implies that a new type of peculiar topological excitation
is generated in the system. Our simulation verifies that the spin defect in
the central region of the trap potential is a giant antiskyrmion with local
topological charge $Q=-2$, and the other six spin defects are six
half-antiskyrmions (antimerons) with respective topological charge $Q=-0.5$
constituting a half-antiskyrmion (antimeron) necklace, as shown in figure
7(d3). From figures 7(d2)-7(d4), we can conclude that the topological
structure in figure 7(d2) is composite giant-antiskyrmion-antimeron
necklaces made of a central giant antiskyrmion and two concentric antimeron
necklaces.

To the best of our knowledge, these new skyrmion excitations observed in the
present system are remarkably different from previously reported skyrmion
structures in other physical systems, such as rotating two-component BECs
with or without SOC (DDI) \cite%
{Zhai,HuH,XuX,ZhouX,Aftalion,WangH,ZhangXF,YangH,ZhangXF2,Fetter2}, rotating
and rapidly quenched spin-1 BECs with SOC in a harmonic trap \cite{LiuCF2},
and\ non-rotating spin-1 BECs with SOC and DDI in a cigar-shaped trap \cite%
{Kato}. In addition, these novel topological excitations (including vortex
excitations and skyrmion excitations) in the present work allows to be
tested and verified in future experiments.

\section{Conclusion}

In summary, we have investigated a rich variety of ground-state phases and
topological defects of rotating spin-1 BECs with SOC and DDI\ in a harmonic
plus quartic trap. The combined effects of SOC, DDI and rotation on the
ground-state configurations and spin textures of the system are analyzed
systematically. For fixed rotation frequency, a ground-state phase diagram
is given with respect to SOC strength and DDI strength. The system can show
rich ground-state phases including vortex string phase with DPs, triangular
(square, pentagonal) vortex lattice phase with DPs, checkerboard phase,
stripe phase with hidden vortices and antivortices, multi-layer vortex
necklaces with a giant vortex and hidden vortex necklaces, depending on the
complex competition between the SOC, DDI and rotation. The increase of SOC
strength or DDI strength or rotation frequency can lead to a series of
structural phase transitions, such as the transition from a square vortex
lattice phase with DPs to a checkerboard phase, the switching from a
checkerboard phase to a stripe phase, and the transition from a pentagonal
vortex lattice phase with DPs to multi-layer vortex necklaces. In addition,
the system favors exotic new skyrmion excitations, such as an antiskyrmion
pair, antiskyrmion-antimeron cluster, skyrmion-antiskyrmion lattice,
skyrmion-antiskyrmion cluster, skyrmion-antiskyrmion-meron-antimeron
lattice, double-layer half-antiskyrmion necklaces, and composite
giant-antiskyrmion-antimeron necklaces. It is shown that for small rotation
frequency, the SOC effect and DDI effect play a key role in determining the
ground phase of the system, while for large rotation frequency, the rotation
effect is dominant. Furthermore, as three important degrees of freedom, the
DDI, SOC, and rotation can be used to achieve the desired ground-state
phases and to control the phase transition between various ground states of
spin-1 BECs in an anharmonic trap. As the ground states are stable against
perturbation and have longer lifetime in contrast to the other stationary
states of the system, we expect that the ground-state structures and
topological defects of the system can be tested and observed in the future
cold atom experiments. The findings in the present work have enriched our
new understanding for topological excitations in cold atom physics and
condensed matter physics.

\begin{acknowledgments}
This work was partially carried out during a visit of the corresponding
author (LW) to the research group of Professor W. Vincent Liu at The
University of Pittsburgh. The authors are extremely grateful to anonymous
reviewers for their careful reading of the manuscript, insightful comments
and constructive suggestions. This work was supported by the National
Natural Science Foundation of China (Grant Nos. 11475144 and 11047033), the
Natural Science Foundation of Hebei Province (Grant Nos. A2019203049 and
A2015203037), and Research Foundation of Yanshan University (Grant No. B846).
\end{acknowledgments}

\end{document}